\documentclass[aps,pra,showpacs,groupedaddress,superscriptaddress,twocolumn]{revtex4-1}
\usepackage{amssymb}
\usepackage{graphicx}
\usepackage{amsmath}
\usepackage{subfigure}
\usepackage{float}
\usepackage{multirow}
\usepackage{mathrsfs}

\begin{document}

\title{Dispersive Jaynes-Cummings Hamiltonian describing a two-level atom
interacting with a two-level single mode field}
\author{C.J.S. Ferreira}
\email{cristinojsf66@gmail.com}
\affiliation{Departamento de F\'{i}sica, Universidade Federal da Para\'{i}ba,  58.051-970 Jo\~{a}o
Pessoa, PB, Brazil}

\author{C. Valverde}
\affiliation{Unidade de Ci\^{e}ncias Exatas e Tecnol\'{o}gicas, Universidade
Estadual de Goi\'{a}s, BR 153, km 98, 75001-970 An\'{a}polis, GO, Brazil} 
\affiliation{Universidade Paulista, BR 153, km 7, 74845-090 Goi\^{a}nia, GO,
Brazil}

\author{B. Baseia}
\affiliation{Departamento de F\'{i}sica, Universidade Federal da Para\'{i}ba,  58.051-970 Jo\~{a}o
Pessoa, PB, Brazil} 
\affiliation{Instituto de F\'{i}sica, Universidade Federal de Goi\'{a}s,
74.690-900 Goi\^{a}nia, GO, Brazil}
\date{\today }

\begin{abstract}
We investigate the time evolution of statistical properties of a single mode
radiation field after its interaction with a two-level atom. The entire
system is described by a dispersive Jaynes-Cummings Hamiltonian assuming the
atomic state evolving from an initial superposition of its excited and
ground states, $\vert e\rangle +\vert g\rangle ,$ and the field evolving
from an initial superposition of two excited levels, $\vert n_{1}\rangle+
\vert n_{2}\rangle$. It is found that the field evolution is periodic, the
period depending on the ratio $n_{2}/n_{1}.$ The energy excitation
oscillates between these two states and the statististics can be either sub-
or super-Poissonian, depending on the values $n_{1},$ $n_{2}$.
\end{abstract}


\maketitle

\section{Introduction}

More than 50 years ago a very simple model Hamiltonian was proposed by E.T.
Jaynes and F.W. Cummings (JC) to study the interaction between a two-level
atom and a quantized single mode field \cite{Jaynes}. The aim of the authors
was to compare the quantum theory with the semi-classical one for the
radiation field. Initially, no difference was observed, either using a
classical field $E(t)=E_{0}cos(\omega t)$ or a quantized field assumed in
one of the most nonclassical state, as the number state $|n\rangle $. Next,
17 years later a new calculation by Eberly et al in 1980 \cite{Eberly} was
implemented assuming the field initially in a coherent state; they showed
the atom exhibiting a new nonclassical effect, then named as
\textquotedblleft collapse and revival\textquotedblright\ of the atomic
inversion. The effect was observed experimentally in 1987 by G. Rempe et. al 
\cite{Rempe}.

The JC model is written in the form ($\hbar =1$), 
\begin{equation}
\mathcal{\hat{H}}=\omega \hat{a}^{\dagger }\hat{a}+\frac{1}{2}\omega _{0}%
\hat{\sigma}_{z}+\lambda (\hat{\sigma}_{+}\hat{a}+\hat{a}^{\dagger }\hat{%
\sigma}_{-}).  \label{e1}
\end{equation}%
In the Eq.(\ref{e1}) $\hat{a}$ ($\hat{a}^{\dag }$) stands for the
annihilation (creation) operator, $\sigma _{-}$ ($\sigma _{+}$) is the
lowering (raising) operator, $\hat{n}=\hat{a}^{\dag }\hat{a}$ is the number
operator, $\omega $ $(\omega _{0})$ is the field (atomic) frequency, and $%
\lambda $ stands for the atom-field coupling. One identifies $\mathcal{\hat{H%
}}_{0}=\omega \hat{a}^{\dagger }\hat{a}+\frac{1}{2}\omega _{0}\hat{\sigma}%
_{z}$\ as the \textquotedblleft free Hamiltonian\textquotedblright\ whereas $%
\hat{V}=$ $\lambda (\hat{\sigma}_{+}\hat{a}+\hat{a}^{\dagger }\hat{\sigma}%
_{-})$\ is the \textquotedblleft interaction Hamiltonian\textquotedblright .
When $\omega =$ $\omega _{0}$ the atom and the radiation field are resonant
and we have $[\mathcal{\hat{H}}_{0},\hat{V}]=0$; as consequence the
interaction $\hat{V}$ in the interaction picture is the same, namely, $\hat{V%
}^{I}=\hat{V}$, and this makes easier the subsequent calculations. Now, when 
$\Delta \omega =\omega _{0}-\omega \neq 0$ the atom and the field are no
longer in resonance; in this case two alternatives may occur: they can be
either near the resonance, which means $\frac{\Delta \omega }{\lambda }%
\approx 1,$ or far from resonance, $\frac{\Delta \omega }{\lambda }\gg 1$.
In the first case we sum and subtract $(\omega \hat{\sigma}_{z}/2)$ to the
resonant case and obtain,

\begin{equation}
\mathcal{\hat{H}}^{\prime }=\omega (\hat{a}^{\dagger }\hat{a}+\frac{1}{2}%
\hat{\sigma}_{z})+\frac{1}{2}\Delta \omega \hat{\sigma}_{z}+\lambda (\hat{%
\sigma}_{+}\hat{a}+\hat{a}^{\dagger }\hat{\sigma}_{-}).  \label{e2}
\end{equation}%
In Eq.(\ref{e2}), the following changes took place:\newline
the new \textquotedblleft free Hamiltonian\textquotedblright\ is $\mathcal{%
\hat{H}}_{0}^{\prime }=\omega (\hat{a}^{\dagger }\hat{a}+\frac{1}{2}\hat{%
\sigma}_{z})$ whereas $\hat{V}^{\prime }=\frac{1}{2}\Delta \omega \hat{\sigma%
}_{z}+\lambda (\hat{\sigma}_{+}\hat{a}+\hat{\sigma}_{-}\hat{a}^{\dag })$\
stands for the new and effective interaction. Then, while $[\mathcal{\hat{H}}%
_{0},\hat{V}]\neq 0$\ we have that $[\mathcal{\hat{H}}_{0}^{\prime },\hat{V}%
^{\prime }]=0$ and, as consequence, the subsequent calculations become
easier again, since $\hat{V}^{\prime I}=\hat{V}^{\prime }$. In both previous
cases (in-resonance and off-resonance) we get exact solutions.

Now, when the detuning $\Delta \omega $ is large, namely: $\frac{\Delta
\omega }{\lambda }\gg 1$, the mentioned easiness to solve the problem no
longer occurs. In this case a good approximation can be obtained from an
equivalent\ interaction Hamiltonian $\hat{V}^{\prime \prime }$, constructed
from the Eq.(\ref{e2}), named dispersive JC model, given by

\begin{equation}
\hat{V}^{\prime \prime }=\lambda ^{\prime }\hat{n}(|i\rangle \langle
i|+|e\rangle \langle e|),  \label{e3}
\end{equation}%
where $\lambda ^{\prime }$ is the effective coupling involving the parameter 
$\lambda $ and the frequency shift $\Delta \omega $\ of the cavity (detuning 
$\Delta \omega \gg \lambda $).

\section{Dispersive JC Model}

The Eq.(\ref{e3}) represents the dispersive JC model: it describes the field
interacting with a 2-level atom, with respect to the levels $|e\rangle $ and 
$|i\rangle $, where $|i\rangle $ represents a virtual state. Hence, when
initially an atom previously prepared in a superposed state $|\psi
_{A}\rangle =(|e\rangle +|g\rangle )/\sqrt{2}$ \cite{Nota} enters a
microwave cavity that contains a field in an arbitrary normalized state $%
|\psi _{F}\rangle ,$ there is no exchange of energy between the atom and the
field with respect to the atomic levels $|e\rangle $ and $|g\rangle $.

In this scenario, the unitary operator $\hat{U}(t)=\exp (-it\hat{V}^{\prime
\prime })=\exp [-it\lambda ^{\prime }\hat{n}(|i\rangle \langle i|+|e\rangle
\langle e|)]$ describes in which way the atom-field system evolves in time.
Thus we have, 
\begin{equation}
|\psi _{AF}(t)\rangle =\frac{1}{\sqrt{2}}\hat{U}(t)[(|e\rangle +|g\rangle
)|\psi _{F}\rangle ].  \label{e4}
\end{equation}%
Next, after using some algebra involving the equality $(\hat{V}^{\prime
\prime })^{n}=\omega ^{n}\hat{n}^{n}(|i\rangle \langle i|+(-1)^{n}|e\rangle
\langle e|)$ we obtain the \textit{entangled} state, with $\phi (t)=\lambda
^{\prime }t,$ 
\begin{equation}
|\psi _{AF}(t)\rangle =|g\rangle |\psi _{F}\rangle +|e\rangle(e^{i\phi (t)%
\hat{n}}|\psi _{F}\rangle ).  \label{e5}
\end{equation}

The Eq.(\ref{e5}) shows that the dispersive interaction affects only the
phase of the field, thus no exchange of energy occurs between the atom and
the field. Also, the Eq.(\ref{e5}) corresponds to the state of the combined
atom-field system right after the atom has crossed the cavity, at an intant
of time $t=\tau $. The evolution of the field state occurs during the
atom-field interaction in the time interval $0\leq t\leq \tau $; $\tau $ is
defined by the speed of the atom and the length of the cavity.

The JC model in these three versions: resonant, near resonance, and far from
resonance, has been explored by many researchers. As examples of works in
the first scenario we mention the Refs. \cite{Simon, Meystre}; in the second
scenario \cite{Scully, Knight} and in the third \cite{Bertet, Peixoto}. In
the latter case, the approach became important to treat the Schr\"{o}dinger
cat problem \cite{Haroche1, Kuklinski}. It is worth mentioning that the JC
model may include various extensions: one of them adds the counter rotating
term $\lambda ^{\prime }(\hat{a}^{\dagger }\hat{\sigma}_{+}+\hat{a}\hat{%
\sigma}_{-})$ to the usual interaction \cite{Walls}; it comes from the
original interaction, $\hat{V}=\mathbf{\hat{d}}\cdot \mathbf{\hat{E},}$
where $\mathbf{\hat{d}\sim (}\hat{\sigma}_{+}+\hat{\sigma}_{-})$ and $%
\mathbf{\hat{E}\sim (}\hat{a}^{\dagger }+\hat{a})$; another extension treats
the case of multiphoton interaction $\lambda \lbrack \hat{a}^{p}\hat{\sigma}%
_{+}+(\hat{a}^{\dagger })^{p}\hat{\sigma}_{-}],$ $p=2,3,4,...$ \cite{Singh,
Chaba, Gerry}; etc. Other kinds of extensions also appear in the literature:
one of them extends the JC model to another, named Buck-Sukumar model \cite%
{Buck}; another interpolates between the JC model to the Sivakumar model 
\cite{Siva}; yet another going from the JC model to the Rodr\'{\i}guez--Lara
model \cite{Lara}; a model that includes `\textit{quonic}' particles was
also proposed: it extends the JC Model to the Shanta-Chaturvedi-Arinivasan
model \cite{Shanta}. A generalized model (VB model) that includes all these
previous (\textit{bosonic}) models has been proposed \cite{Valverde}.
Although not very usual, the JC model is also treated in the Heisenberg
picture \cite{Stenholm}. Now, according to the Eq.(\ref{e5}),if we let the
traveling atom traverse a second Ramsey zone \cite{Haroche2}, this apparatus
leads the atom in the state $|e\rangle $ to the superposed state $(|e\rangle
+|g\rangle )$ and leads the atom in the state $|g\rangle $ to $(|g\rangle
-|e\rangle )$. Then we get, 
\begin{equation}
|\psi _{AF}(t)\rangle =|g\rangle (|\psi _{F}\rangle +|\psi _{F}^{\prime
}\rangle )+|e\rangle (|\psi _{F}^{\prime }\rangle -|\psi _{F}\rangle ),
\label{e6}
\end{equation}%
\newline
where $|\psi _{F}^{\prime }\rangle =e^{i\phi \hat{n}}|\psi _{F}\rangle $.
Next, if the atom is detected in its ground state $|g\rangle $, the field
inside cavity is projected in the even superposition $|\psi _{F}(t)\rangle
=|\psi _{F}\rangle +e^{i\phi \hat{n}}|\psi _{F}\rangle $. As an application
we assume the field state initially in the normalized superposition of two
excited number state \cite{malboui}: $|\psi _{F}(0)\rangle =(|n_{1}\rangle
+|n_{2}\rangle )/\sqrt{2}$. Substituting this state in the Eq.(\ref{e6}) we
obtain the evolved field in the state, with $\phi =\phi (t)=\lambda ^{\prime
}t$, 
\begin{eqnarray}
|\psi _{F}(t)\rangle  &=&\frac{1}{\sqrt{2}}[(|n_{1}\rangle +|n_{2}\rangle
)+(e^{i\phi \hat{n}_{1}}|n_{1}\rangle +e^{i\phi \hat{n}_{2}}|n_{2}\rangle )]
\notag  \label{e7} \\
&=&\frac{2}{\sqrt{2}}[e^{\frac{i\phi n_{1}}{2}}(\cos \frac{\phi n_{1}}{2}%
)|n_{1}\rangle +e^{\frac{i\phi n_{2}}{2}}(\cos \frac{\phi n_{2}}{2}%
)|n_{2}\rangle ].  \notag \\
&&
\end{eqnarray}%
We now can write the normalized state as 
\begin{equation}
|\psi _{F}(t)\rangle =\eta \lbrack e^{\frac{i\phi n_{1}}{2}}(\cos \frac{\phi
n_{1}}{2})|n_{1}\rangle +e^{\frac{i\phi n_{2}}{2}}(\cos \frac{\phi n_{2}}{2}%
)|n_{2}\rangle ],  \label{e8}
\end{equation}%
where $\eta =(\cos ^{2}\theta _{1}+\cos ^{2}\theta _{2})^{-1/2}$ is the
normalization factor, with $\theta _{i}=\phi n_{i}/2,$ $i=1,2.$

\section{Statistical Properties}

The result in Eq.(\ref{e8}) allows us to obtain statistical properties of
the field, as follows:

\subsection{Statistical Distribution}

From the Eq.(\ref{e8}) we obtain the time evolution of the statistical
distribution $P_{n}(t)$. We have,

\begin{eqnarray}
P_{n}(t) &=&|\langle n|\psi _{F}(t)\rangle |^{2}  \notag \\
&=&\eta ^{2}|e^{\frac{i\phi n_{1}}{2}}(\cos \frac{\phi n_{1}}{2})\delta
_{n,n_{1}}+e^{\frac{i\phi n_{2}}{2}}(\cos \frac{\phi n_{2}}{2})\delta
_{n,n_{2}}|^{2}.  \notag \\
&&
\end{eqnarray}%
%
%
%
%
%
%
%
%
%
%
%

\begin{figure}[]
\includegraphics[scale=0.465]{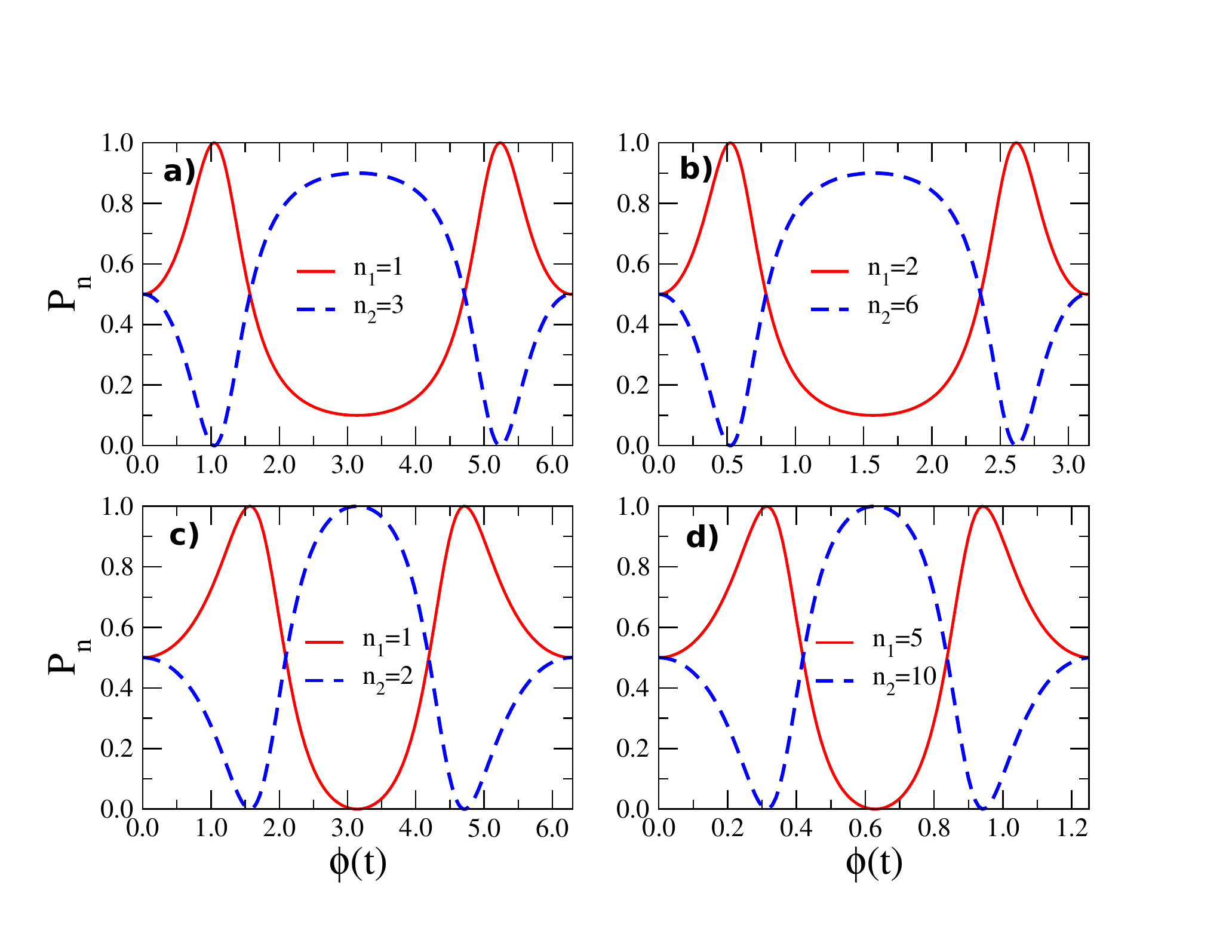}
\caption{Time evolution of statistical distribution for the two field
components $|n_1\rangle$ and $|n_2\rangle$, it shows in which way these
components exchange their energies: (a) for $n_{1}=1$ and $n_{2}=3$; (b) for 
$n_{1}=2$ and $n_{2}=6$; (c) for $n_{1}=1$ and $n_{2}=2$; (d) for $n_{1}=5$
and $n_{2}=10$.}
\label{f1}
\end{figure}

Since in this case we find that $P_{n}(t)=1$ for all times, then the
relevant behavior concerns the statistics $P_{n_{1}}(t)$ and $P_{n_{2}}(t):$
they show in which way the exchange of excitations occurs between the
components $|n_{1}\rangle $ and $|n_{2}\rangle $. Figs.\ref{f1}(a), \ref{f1}%
(b), \ref{f1}(c) and \ref{f1}(d) show the time evolution of the statistical
distributions $P_{n_{1}}(t)$ and $P_{n_{2}}(t)$ for the field in the state $%
|\psi _{F}(t)\rangle $, for various values of the excitations $n_{1}$ and $%
n_{2}$. The behaviors of these distributions are oscillatory, e.g., with
period $\tau =2\pi $ for the pairs ($n_{1}=1$, $n_{2}=3$) and ($n_{1}=1$, $%
n_{2}=2$) and $\tau =\pi $ for the pair ($n_{1}=2$, $n_{2}=6$) and $\tau
=2\pi /5$ for the pair ($n_{1}=5$, $n_{2}=10$). We note that, although
having different periods, all pairs $(n_{1},n_{2})$ and $(n_{1}^{\prime
},\,n_{2}^{\prime })$ have the same behavior when $n_{1}^{\prime
}/n_{1}=n_{2}^{\prime }/n_{2}=p$, with $p=2,\,3,\,4,...\,$In addition we
observe that when $P(n_{i})=1$ all field excitation concentrates into the
component $|n_{i}\rangle $, as it should. For example, the state $%
|n_{1}\rangle =|1\rangle $ becomes pure at $\tau ^{\prime }=1.05$ and $\tau
^{\prime \prime }=5.24$ (in Fig.\ref{f1}(a)) and at $\tau ^{\prime }=\pi /2$
and $\tau ^{\prime \prime }=3\pi /2$ (in Fig.\ref{f1}(c)). At these points
the entire field state becomes a number state, $|\psi _{F}(t)\rangle
=|n_{1}\rangle .$ 

\subsection{Mandel Parameter}

\begin{figure}[]
\includegraphics[scale=0.0605]{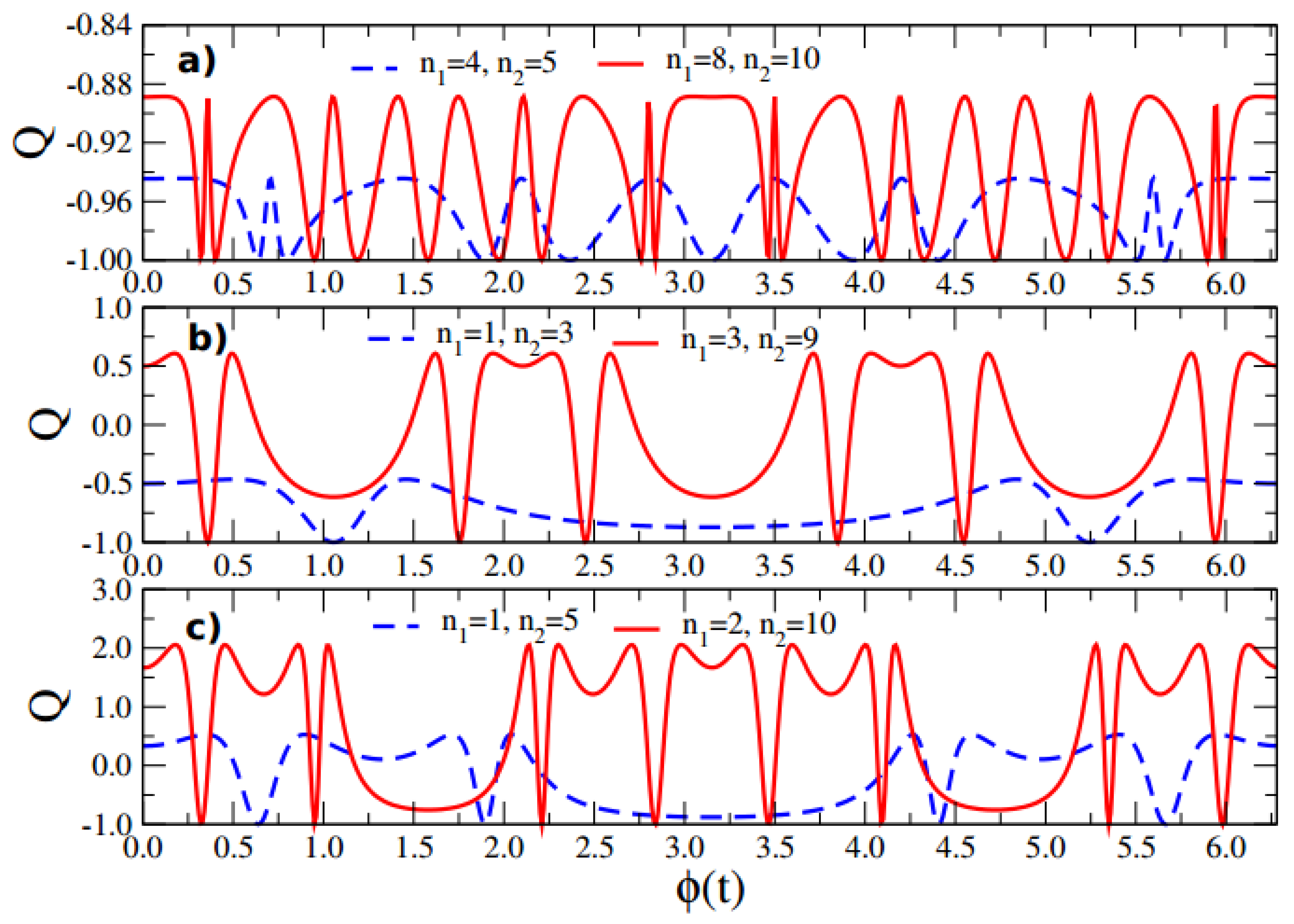}
\caption{(a)-Plots for the Mandel parameters for the pairs $(n_{1},
n_{2})=(4, 5)$ and $(8, 10)$ respectively; (b)- Same as in Fig.\protect\ref%
{f2}(a), for $(n_{1}, n_{2})=(1, 3)$ and $(3, 9)$; (c) Same as in Fig.%
\protect\ref{f2}(a), for $(n_{1}, n_{2})=(1, 5)$ and $(2, 10)$.}
\label{f2}
\end{figure}

The Mandel parameter informs whether the statistics is either Poissonian,
sub-Poissonian or super-Poissonian \cite{Mandel}. To this end we must first
calculate the variance of photon number, $\langle (\triangle \hat{n}%
)^{2}\rangle =\langle \hat{n}^{2}\rangle -\langle \hat{n}\rangle ^{2}.$ From
the Eq.(\ref{e8}) we obtain, for $\langle \hat{n}(\phi )\rangle $, $\langle 
\hat{n}^{2}(\phi )\rangle $ and $\langle \hat{n}(\phi )\rangle ^{2}$, with $%
\phi =\phi (t)=\lambda ^{\prime }t$, 
\begin{equation}
\langle \hat{n}(\phi )\rangle =\eta ^{2}[n_{1}\cos ^{2}(\frac{\phi n_{1}}{2}%
)+n_{2}\cos ^{2}(\frac{\phi n_{2}}{2})],  \label{e9}
\end{equation}%
\begin{equation}
\langle \hat{n}^{2}(\phi )\rangle =\eta ^{2}[n_{1}^{2}\cos ^{2}(\frac{\phi
n_{1}}{2})+n_{2}^{2}\cos ^{2}(\frac{\phi n_{2}}{2})],  \label{e10}
\end{equation}%
and $\langle \hat{n}(\phi )\rangle ^{2}$ is obtained from the Eq.(\ref{e9}).
Thus, from Eqs.(\ref{e9}) and (\ref{e10}) we obtain the Mandel parameter. 
\begin{eqnarray}
Q &=&\frac{\langle (\Delta \hat{n})^{2}\rangle -\langle \hat{n}\rangle }{%
\langle \hat{n}\rangle }  \notag \\
&=&\frac{1}{\langle \hat{n}\rangle }\{\eta ^{2}[n_{1}^{2}\cos ^{2}(\frac{%
\phi n_{1}}{2})+n_{2}^{2}\cos ^{2}(\frac{\phi n_{2}}{2})]  \notag \\
&&-\eta ^{4}[n_{1}^{2}\cos ^{4}(\frac{\phi n_{1}}{2})+n_{2}^{2}\cos ^{4}(%
\frac{\phi n_{2}}{2})  \notag \\
&&+2n_{1}n_{2}\cos ^{2}(\frac{\phi n_{1}}{2})\cos ^{2}(\frac{\phi n_{2}}{2}%
)]-\langle \hat{n}\rangle \}.  \label{e11}
\end{eqnarray}
Figs.\ref{f2}(a), \ref{f2}(b) and \ref{f2}(c) show the evolution of the
Mandel parameter for various excitation values of the initial excitations in
the components $|n_{1}\rangle$ and $|n_{2}\rangle$. 
All plots in Fig.\ref{f2}(a) exhibit the sub-Poissonian effect ($-1<Q<0$)
during their respective periods $\tau=2\pi$ and $\tau=\pi$ for the pairs $%
(n_{1},n_{2})$: $(4, 5)$ and $(8, 10)$, respectively. In Fig.\ref{f2}(b) the
pair $(1, 3)$ exhibit sub-Poissonian effect during all period, however, the
pair ($3, 9$) exhibit sub- and super-Poissonian effect during the period $%
\tau=2\pi/3$. The sub-Poissonian effect is shown for very short time
intervals in Fig.\ref{f2}(c); Concerning the Fig.\ref{f2}(c), the state
remains sub-Poissonian for times in the respective periods, as also shown in
Fig.\ref{f2}(b).

Now, all results obtained above can be extended to initial field states with
two Fock components having unequal weights, e.g., $|\psi _{F}(0)\rangle
=c_{1}|n_{1}\rangle +c_{2}|n_{2}\rangle $ with $|c_{1}|^{2}+$\ $|c_{2}|^{2}=1
$. At this point a few words should be devoted on how to get an available
initial superposed state of the type used here, as $|\psi _{F}(0)\rangle
=(|n_{1}\rangle +|n_{2}\rangle )/\sqrt{2}$ or one of its extensions. In Fig.(%
\ref{f3}) below, the Fig.\ref{f3}(a) represents a coherent state $|\alpha
\rangle $ inside a good cavity. As well known, by making a conveniently
prepared two-level atom that crosses the cavity and interacts with a
coherent state, it transforms the coherent state into a superposition of two
coherent states, including the so called \textquotedblleft
Schrodinger-cat\textquotedblright\ state when a rotation by an angle $\theta
=\pi $ in the phase space affects the coherent state \cite{Haroche1}. 
\begin{figure}[tbp]
\includegraphics[scale=0.0605]{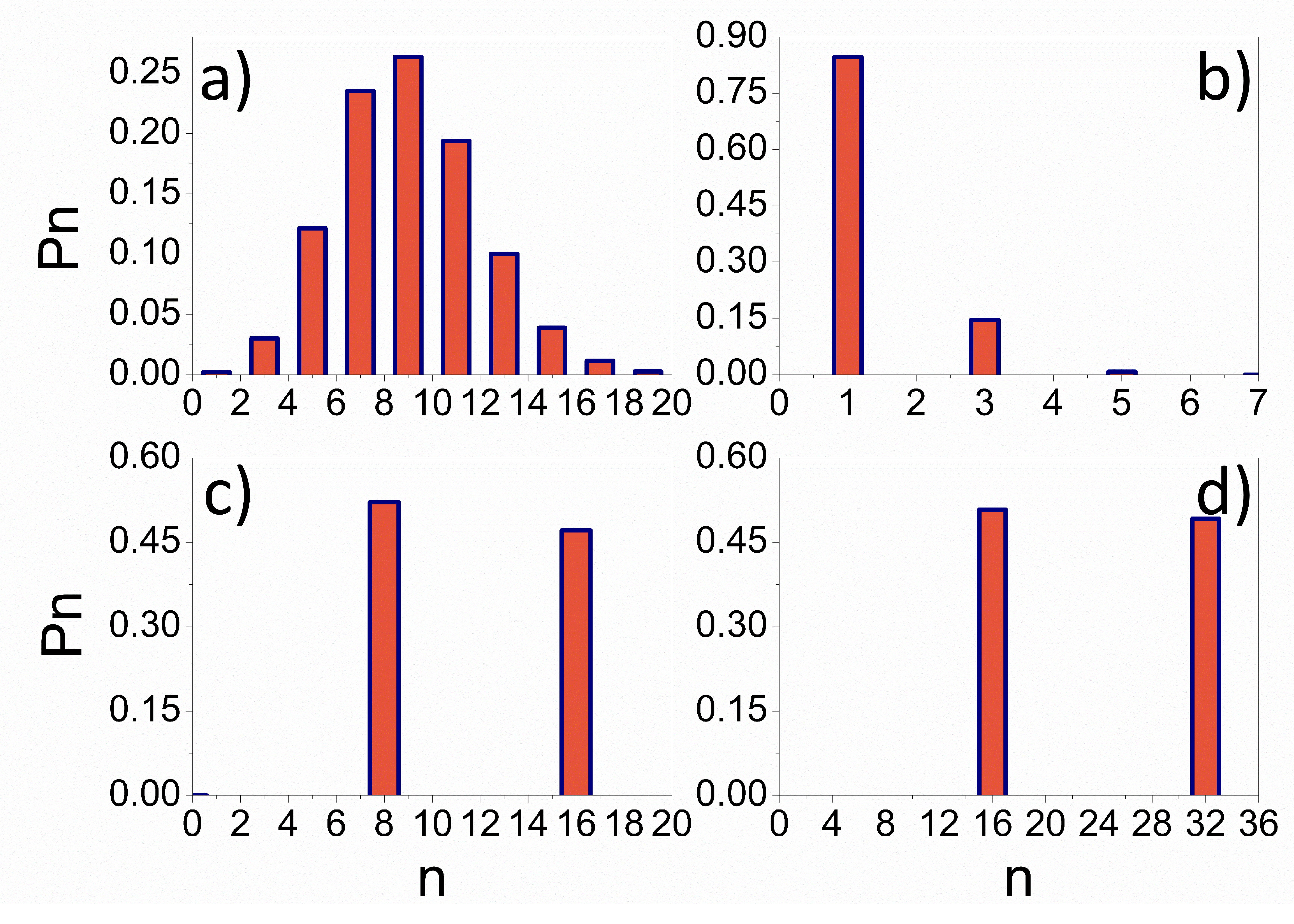}
\caption{Photon number distribition displaying the creation of the
superposition of two number states. a) $\protect\alpha =3.0$ and $N=2;$ b) $%
\protect\alpha =1.009$ and $N=2;$ c)\ $\protect\alpha =3.482$ and $N=3;$ d) $%
\protect\alpha =4.899$ and $N=4.$}
\label{f3}
\end{figure}
Now, as one example, the Ref.\cite{malboui} studied the generation of
various superposition states when $N$ atoms cross the cavity, $N=1,2,3,...$
suscessively with convenient speeds $\upsilon _{N}=\upsilon _{N-1}/2.$ The
wavefunction describing the system is given by, 
\begin{equation}
\left\vert \Psi _{N}^{\pm }(\alpha )\right\rangle =\sum\limits_{n=0}^{\infty
}C_{N}^{\pm }(n;\alpha )\left\vert n\right\rangle ,
\end{equation}%
where, 
\begin{eqnarray}
C_{N}^{\pm }(n;\alpha ) &=&\left\langle n|\Psi _{N}^{\pm }(\alpha
)\right\rangle ,  \notag \\
&=&[2^{2N}\exp (\alpha ^{2})\beta _{N}^{\pm }(\alpha ^{2})]^{-1/2}\exp
(-\left\vert \alpha \right\vert ^{2}/2)\alpha ^{n}  \notag \\
&&\times \frac{(-1)^{n}\pm 1}{\sqrt{n!}}\sum\limits_{j=0}^{2^{N-1}-1}\exp (%
\frac{in\pi j}{2^{N-1}}),
\end{eqnarray}%
with $\beta _{N}^{\pm }(\alpha ^{2})$ for $N>1,$ standing for, 
\begin{eqnarray}
\beta _{N}^{\pm }(\alpha ^{2}) &=&\frac{1}{2^{N-1}}\beta _{1}^{\pm }(\alpha
^{2})+\frac{1}{2^{2(N-1)}}\sum\limits_{k=0}^{2^{N-1}-1}\{(2^{N}-2k)  \notag
\\
&&\times \cos [\alpha ^{2}\sin (\pi k/2^{N-1})]\beta _{1}^{\pm }[\alpha
^{2}\cos (\pi k/2^{N-1})]\},  \notag \\
&&
\end{eqnarray}%
and $\beta _{1}^{\pm }(\alpha ^{2})=\frac{1}{2}[\exp (\alpha ^{2})\pm \exp
(-\alpha ^{2})]$ \ for $N=1$. Thus the probability of photon number
distribution is obtained from the expression, 
\begin{equation}
P_{N}^{\pm }(n;\alpha )=\left\vert \left\langle n|\Psi _{N}^{\pm }(\alpha
)\right\rangle \right\vert ^{2},
\end{equation}%
the sign $(+)$ standing for the even state and sign $(-)$ for the odd state. 

From convenient choices of values of $\alpha $ and $N$ one gets the results
displayed in Fig.\ref{f3}(b), \ref{f3}(c), and \ref{f3}(d). Fig.\ref{f3}(b)
is one of the results when passing two atoms, for the states $|1\rangle $
and $|3\rangle $; Fig.\ref{f3}(c) concerns the case of three atoms, leading
to the states $|8\rangle $ and $|16\rangle $; Fig.\ref{f3}(d) is the case of
four atoms leading to the states $|16\rangle $ and $|32\rangle $. Other
pairs of Fock components can also be obtained, as the approximate pair $%
|4\rangle $ and $|8\rangle $ shown in Fig.4(a) of Ref.\cite{malboui}, using
two atoms.

\section{Conclusion}

The plots of the statistical distributions $P_{n_{1}}$ and $P_{n_{2}}$\ show
in which way the field state $|\psi _{F}(t)\rangle $ shares its excitation
to components $|n_{1}\rangle $ and $|n_{2}\rangle $. We note the
similarities that occur for the pairs of components $|n_{1}\rangle ,$ $%
|n_{2}\rangle $ and $|n_{1}^{\prime }\rangle ,$ $|n_{2}^{\prime }\rangle $
when $n_{1}^{\prime }/n_{1}=n_{2}^{\prime }/n=p,$ $p=1,2,3,...$; they show
the same behavior, but in different periods, e.g., one of them being $\tau $
the other is $\tau ^{\prime }=\tau /p.$ The plots of the Mandel parameter
show the occurrence of sub-Poissonian statistics in Fig.\ref{f2}(a) and
(partial) super-Poissonian statistics in Fig.\ref{f2}(b) and \ref{f2}(c).\
Moreover, we verify that the larger the difference between the values $n_{1}$
and $n_{2}$ $(n_{2}-n_{1}\gg 1)$, the larger is the super-Poissonian
character of the statistics. Again, the same similarities found in Fig.\ref%
{f1} is also observed in Fig.\ref{f2} for the case $n_{1}^{\prime
}/n_{1}=n_{2}^{\prime }/n=p$. During all the state evolution no squeezing
effect was observed and, according to the Ref.\cite{Mandel}, one would
observe this effect only when the field state $|\psi _{F}(t)\rangle $ can
distribute his excitation to more than two Fock components. Finally, some
words were dedicated on how to prepare a generalized initial state $|\psi
_{F}(0)\rangle =c_{1}|n_{1}\rangle +c_{2}|n_{2}\rangle ,$ as one of them
assumed in this report.

\section{Acknowledgements}

We thank the Brazilian funding agencies CAPES, CNPq and FAPEG for partial
supports.


\begin{thebibliography}{99}
\bibitem{Nota} This superposed state $(\vert e\rangle+\vert g\rangle)$ is
obtained by making the atom in its excited state $\vert e\rangle$ crossing a
Ramsey zone.
\end{thebibliography}
\end{document}